\documentclass[12pt,preprint,floats,tightenlines,showpacs]{revtex4}              

\usepackage{amsmath}
\usepackage{amsfonts}
\usepackage{amscd}
\usepackage{epsfig}
\usepackage{amssymb}
\usepackage{tabularx}     

\def\c{{\mathbb C}}
\def\a{\alpha}
\def\b{\beta}
\def\g{\gamma}
\def\e{\epsilon}

\def\d{\delta}
   
\def\k{\kappa}
\def\z{\xi}
\def\w{\omega}

\def\s{\sigma}
\def\p{\partial}

\def\ad{\dot{\alpha}}
\def\bd{\dot{\beta}}
\def\x{{\bf x}}

\begin{document}

\preprint{\vbox{\hbox{hep-th/0302086}}} 

\title{Remarks on experimental  bounds on quantum gravity  effects on
fermions} \author{Gustavo Dotti and Reinaldo J. Gleiser}
\affiliation{FaMAF, Universidad Nacional de C\'ordoba,\\
Ciudad Universitaria, 5000, C\'ordoba, ARGENTINA}
\email{gdotti@fis.uncor.edu}

\begin{abstract}
Effects of space time geometry fluctuations on fermionic fields have recently
been looked for in nuclear physics experiments, and were found to be much lower
than predicted,  at a phenomenological level, by  loop quantum gravity. We show
that possible corrections to the canonical structure in the semi classical
regime may introduce important changes in the outcome of the theory, and may
explain the observed mismatch with experiments.
\end{abstract}

\maketitle

\vspace{.2cm}

\vspace{.2cm}
 
\section{Introduction}

It has only recently been  realized that certain quantum gravity  corrections 
to Maxwell and spinor  field equations can lead to measurable effects. Tiny
corrections  in the propagation of photons or neutrinos  accumulate through
cosmological  distances, giving rise to potentially observable effects
\cite{oe} that would imply a breakdown of Lorentz invariance. In particular, for
photons, this might take the form a birefringence effect \cite{gapu}, but, as
indicated in \cite{glko}, the observations of  polarization in the visible and
ultraviolet light from certain cosmological  sources already imply an important
upper bound on the effect. In the case of fermions, it has been shown in
\cite{suv} that the breakdown  could, in principle,  be confirmed by means of
extremely  sensitive isotropy tests in nuclear systems \cite{suv}.
The results obtained in \cite{glko} and \cite{suv}, however, provide
experimental bounds that seem to indicate that the Lorentz violation, if
present, is far smaller than suggested by the theoretical predictions, pointing
to an apparent discrepancy between theory and experiment. Seeking an
explanation for this mismatch in the case of \cite{suv}, we consider here the
possibility that the canonical structure of the effective low energy theory gets
corrections that vanish in the $\ell_P \to 0$ limit.  This introduces additional
terms in the field equations that may cancel  some of the
Lorentz violating effects,  and therefore conciliate theory with experiment.

Let us start by recalling the action for a Majorana spinor in Minkowski space time
\begin{equation} \label{lagr}
S =  \int d^4 x \, \left[ 
i \bar \z_{\ad} \bar \s^{n \ad \a} \p_n \z_{\a} 
- \frac{m}{2} (\z_{\a}\e^{\a\b} \z_{\b}
 - \bar \z_{\ad}  \e^{\ad \bd} \bar \z_{\bd}) \right]
\end{equation}
Here
${\s^0}_{\a \ad} = \bar \s ^{0\ad \a} = {\mathbb I}$ and 
$-\bar \s ^{j \ad \a} = {\s^j}_{\a \ad} = $ Pauli matrices.
Spinor indices are raised and lowered using the second
index of the antisymmetric tensors $\e^{\a \b}$ and  $\e_{\a \b}$ 
($\e^{12}=\e_{21}=1$, and similarly for $\e^{\ad \bd}$.)
(\ref{mops})The equation of motion that follows from (\ref{lagr}) 
\begin{equation} \label{eq}
i {\bar \s}^{n \ad \a} \p_n \z_{\a} + m \e^{\ad \bd} \bar \z _{\bd} = 0
\end{equation}
gives   the $\ell_p=0$ limit of eq.(111) in \cite{alf} ((7) in \cite{amu},(1) in
\cite{suv}). From (\ref{lagr}) we obtain the field $\pi^{\a}$ conjugate
to $\z_{\a}$ 
\begin{equation}\label{cm}
\pi^{\a} =  i \bar \z_{\ad} \bar \s^{0 \ad \a}
\end{equation}
When performing a $3+1$ decomposition to write down the Hamiltonian, the
$SL(2,\c)$ Lorentz symmetry is broken down to an $SU(2)$ rotation subgroup,
under which dotted and undotted spinors are made equivalent by the $SU(2)$
invariant tensor $ {\s^0}_{\a \ad}$ and its  inverse ${\bar \s} ^{0 \ad \a}$.
Defining ${{\s^k} _{\a}}^{\b} = {-\s ^0}_{\a \ad} \bar \s ^{k \ad \b}, k=1,2,3$,
the Hamiltonian obtained from (\ref{lagr}) can be  written as
\begin{equation}
\label{h} H = \int d^3x \left[\pi^{\b}{{\s^k}_{\b}}^{\a}\p_k \z_{\a}
 +  \frac{m}{2} \left( \z_{\b}  \e^{\b \a} \z_{\a} - \pi^{\b} \e_{\b \a}
\pi^{\a}\right)
\right]
\end{equation}
The canonical equal time anti-commutation relations 
\begin{equation} \label{car}
\{ \z_{\a}(x^0,\x), \pi^{\b}(x^0,\x') \} = i {\delta_{\a}}^{\b} \delta(\x-\x')
\end{equation}
together with the Hamiltonian (\ref{h}) allow us to recover the field equations
(\ref{eq}) for the Majorana spinor
\begin{eqnarray} \nonumber
i \dot{\z}_{\g} &=& [ \z_{\g}, H ] = \int d^3x' 
\left( [ \z_{\g}, \pi^{\b}{{\s^k}_{\b}}^{\a}\p_k \z_{\a} ]
 -  \frac{m}{2} [ \z_{\g} , \pi^{\b} \e_{\b \a}
\pi^{\a} ] \right) \\                    
&=& i{{\s^k}_{\g}}^{\a}\p_k \z_{\a}-  im \e_{\g \b} \pi^{\b}
\end{eqnarray}
The generalization of (\ref{h}) to generic backgrounds gives the fermionic piece
$H_F$ of the fermion-gravity Hamiltonian, eq. (3) in \cite{alf}. In the quantum
regime, operator products in $H_F$ are regularized following Thiemann \cite{t}.
A coherent  $| {\cal L},\z \rangle$ state in the Hilbert space ${\cal H}_{grav}
\otimes {\cal H}_{fermion}$ that approaches a flat space for distances $d \gg
{\cal L} \gg \ell_P$, and a smooth fermion field living in it, is postulated in
\cite{alf,amu}. $| {\cal L},\z \rangle$ is assumed to be peaked around a flat
metric and connection, the expectation value of the spatial metric operator
$q_{ab}$ behaving as $\langle  {\cal L},\z | q_{ab} |  {\cal L},\z \rangle =
\d_{ab} + {\cal O} (\ell_p/{\cal L})$, and also peaked around a fermion field
configuration that varies slowly at the scale ${\cal L}$. A low energy effective
Hamiltonian is then defined as the expectation value of the regularized $H_F$ in
this state. The result is eq.(109) in \cite{alf}:
\begin{multline} \label{eh}
H_{eff} \equiv \langle  {\cal L},\z | H_F |  {\cal L},\z \rangle \\
= \int d^3x \left[\pi^{\b}{{\s^k}_{\b}}^{\a} \hat A \p_k \z_{\a}  + 
\frac{i}{2 {\cal L}} \pi^{\a} \hat C \z_{\a} 
 + \frac{m}{2} \left( \z_{\b}  \e^{\b \g} {O_{\g}}^{\a} \z_{\a} - \pi^{\b}
 {O_{\b}}^{\g} \e_{\g \a} \pi^{\a}\right) \right] .
\end{multline}
Here $\pi^{\a} \equiv i \bar \z_{\ad} \bar \s^{0 \ad \a}$ (eq. (\ref{cm})),
${O_{\g}}^{\e} = \a {\d_{\g}}^{\e} -i \b {{\s^k}_{\g}}^{\e} \p_k $ and 
\begin{eqnarray} \nonumber
\hat A &=& 1 + \k_1 \left( \frac{\ell_P}{{\cal L}} \right)^{1+\Upsilon} 
+  \k_2 \left( \frac{\ell_P}{{\cal L}} \right)^{2+2\Upsilon}  
+ \frac{\k_3}{2}  \ell_P ^2 \nabla^2 \\
\hat C &=& \k_4 \left( \frac{\ell_P}{{\cal L}} \right)^{\Upsilon}  +
 \k_5 \left( \frac{\ell_P}{{\cal L}} \right)^{1+2\Upsilon}  +
 \k_6 \left( \frac{\ell_P}{{\cal L}} \right)^{2+3\Upsilon}  +
\frac{\k_7}{2}  \left( \frac{\ell_P}{{\cal L}} \right)^{\Upsilon} {\ell_P}^2
 \nabla^2 \label{conv} \\
\a &=& 1+\k_8  \left( \frac{\ell_P}{{\cal L}} \right)^{1+\Upsilon} 
\hspace{1cm} \b/\ell_P = \frac{\k_9}{2} +  \frac{\k_{11}}{2} 
\left( \frac{\ell_P}{{\cal L}} \right)^{1+\Upsilon} \nonumber
\end{eqnarray}
$\Upsilon$ is a positive constant introduced in \cite{alf} to allow non integer
powers of $(\ell_p/{\cal L})$ in the expansion  of the expectation value of the
connection, a possibility that was not considered in the previous works
\cite{suv,amu}, where $\Upsilon=0$. The dimensionless constants $\k_j$ are
expected to be of order unity. An explicit construction of $| {\cal L}, \z
\rangle$ -which is lacking- would allow us to  evaluate all these constants.
Similar derivations can be found in \cite{gapu} for Maxwell fields and \cite{t2}
for scalar, Maxwell and fermion fields, with the following  conceptual
difference: the effective Hamiltonian is defined as the {\em partial}
expectation value $H_{eff} \equiv \langle  {\cal L} | H_F |  {\cal L} \rangle$
in the gravity sector ${\cal H}_{grav}$. Since the regularized fermionic
Hamiltonian is normal ordered \cite{t2} and $| {\cal L}, \z \rangle$ coherent in
the fermionic sector, both results look formally equal. However, in Thieman's
approach, (\ref{eh}) is an {\em operator} in ${\cal H}_{fermion}$. From
(\ref{car}) and (\ref{eh}) we obtain the field  equation eq.(111) in \cite{alf}
((7) in \cite{amu},(1) in \cite{suv})
\begin{equation} \label{fe}
\left[i \partial_t -i \hat A \s^j \p_j + \frac{\hat C}{2 {\cal L}} \right] \z 
+ m \left( \a -i \b \s ^j \p_j \right) i\s^2 \z^* = 0, 
\end{equation}
which reproduces (\ref{eq}) in the limit $\ell_P \to 0$, and gives  QG
corrections up to order $(E \ell_P)^2$, $E$ a characteristic energy scale for
the fermion \cite{amu,suv}. \\ Equation (\ref{fe}) with  $\Upsilon=\k_4=0$, and
keeping only leading order corrections (i.e.,  setting all $\k_j=0$ except for
$j=1,5,9$), was used in \cite{suv} to obtain a modified Dirac equation depending
on $\k_1, \k_5$ and $ \k_9$. This modified Dirac equation violates
Lorentz symmetry, and therefore gives the time evolution in a preferred 
frame, which is understood to be the CMB frame. The equation actually follows 
from the lagrangian \cite{suv}
\begin{multline} \label{dirac}
L_D = \frac{i}{2}  \bar \Psi \g^a \p_a \Psi - \frac{m}{2}
 \bar \Psi  \Psi + \frac{i}{2} \k_1 (m \ell_p) \bar \Psi
\g_a (g^{ab} - W^a W^b) \p_b \Psi \\
+ \frac{1}{8} \k_9 (m \ell_p) \bar \Psi \e_{abcd} W^a \g^b \g^c \p^d  \Psi 
- \frac{1}{4}  \k_5  (m \ell_p) m W_a \bar \Psi \g_5 \g^ a \Psi 
\end{multline}   
if we set  $W^a=(1,0,0,0)$. Therefore (\ref{dirac})  generalizes the theory  to
other frames if   $W^a$ is replaced by the measured CMB frame's four-velocity
($|\vec W | \simeq 1.23 \times  10^{-3} c$ from the Earth.) Lorentz violation
comes entirely from the fixed 4-vector $W^a$ in (\ref{dirac}). However, this
violation is severely restricted by data on high precision tests of rotational
symmetry  in atomic and nuclear  systems, which, as shown in \cite{suv}, can be
used to set the following stringent bounds on the constants $\k_1,\k_5$ and
$\k_9$, expected in principle to be of order unity:
\begin{equation} \label{bounds} \left| \k_1  \right| < 3 \times 10^{-5}
\hspace{1cm} \left| \k_9 +  \k_5  \right| < 4 \times 10^{-9}
\end{equation}
The above bounds suggest that theory and experimental evidence will only 
agree if, after constructing the semiclassical states and fulfilling the details
left over in the derivation of (\ref{fe}),  one finds that  $\k_1=0$ and
$\k_9=-\k_5$, i.e., the first Lorentz violating term in (\ref{dirac}) is absent,
and the other two appear with suitable coefficients. While cancellations of
terms of the same order cannot be excluded, there appears to be no particular
reason for a small $\k_1$, and its smallness is particularly intriguing
\cite{suv}. In the next Section we suggest an alternative formulation for the
{\em effective low energy} description of quantum gravity in the femion sector.
It makes essential use of the possibility that the choice of appropriate
canonical variables for the effective theory may require the inclusion of
corrective terms in the anticommutation relations, in such a way one may
recover agreement with the observational bounds in a more natural way. The
effect of these terms on the phenomenological description of neutrino
propagation and some low energy nuclear physics experiments are considered
respectively in Sections 3 and 4. The last Section is devoted to a brief
summary and conclusions.

\section{An extension of the formalism}

Looking for an alternative explanation, we allow non integer powers of
$(\ell_p/{\cal L})$ by restoring  $\Upsilon$ in the operators (\ref{conv}), and
further  consider the possibility that, in the effective theory, $\z$ and
$ i \bar \z_{\ad} \bar \s^{0 \ad \a}$ do not anticommute canonically.
This possibility arises quite naturally in an effective lagrangian approach.
If we replace   $\pi^{\a} \to  i \bar \z_{\ad} \bar \s^{0 \ad \a}$ back 
in the expectation value  (\ref{eh}) we obtain the effective energy 
\begin{multline} \label{ee}
\int d^3x \left[ i \bar \z_{\bd} \bar \s^{0 \bd \b}{{\s^k}_{\b}}^{\a} \hat A
 \p_k \z_{\a}+\frac{i}{2 {\cal L}}  i \bar \z_{\ad} \bar \s^{0 \ad \a} \hat C
\z_{\a}  + \frac{m}{2} \left( \z_{\b}  \e^{\b \g} {O_{\g}}^{\a} \z_{\a}
-  i \bar \z_{\bd} \bar \s^{0 \bd \b} {O_{\b}}^{\g} \e_{\g \a} 
 i \bar \z_{\ad} \bar \s^{0 \ad \a} \right)\right] \\
\equiv \int d^3x \, {\cal E}_{eff} \equiv E_{eff}.
\end{multline}
Dropping for simplicity, as in \cite{suv}, sub leading order corrections in 
(\ref{conv}) (in particular, terms containing  Laplacians) yields ${\cal
E}_{eff} = {\cal E}_{eff}(\z,\bar \z,\vec{\nabla} \z,\vec{\nabla} \bar \z)$.
There is, however, a large freedom in the construction of effective lagrangians
${\cal L}_{eff}$ leading to the  energy density ${\cal E}_{eff}$. In fact, since
${\cal L}_{eff}$  is expected to be linear in $\dot{\z}$, for any
$f^{\a}(\z,\bar \z,\vec{\nabla} \z, \vec{\nabla} \bar \z)$,
\begin{equation} \label{el}
{\cal L}_{eff} = \{ \dot \z _{\a} f^{\a}(\z,\bar \z, \vec{\nabla}
\z,\vec{\nabla} \bar \z) +cc \}
- {\cal E}_{eff}(\z,\bar \z,\vec{\nabla} \z,\vec{\nabla} \bar \z)
\end{equation}
will do the job, since we  loose track of the conjugate  field $f^{\a}$ in  the
energy density. Different choices of $ f^{\a}$ will, however, give different
physics. Back to the Hamiltonian formalism, the dynamical information is
recovered through the  equal time anti-commutation relations
\begin{equation} \label{c}
\{\z_{\a},f^{\b}(\bar \z)\} = i {\d_{\a}}^{\b} \d(\x-\x').
\end{equation}
together with  the effective energy (\ref{ee}). Equivalently, we may adopt
(\ref{cm}) as a  {\em definition} and use (\ref{eh}), keeping in mind that
$\pi$  and $\z$ {\em do not} anticommute canonically. In view of (\ref{c}), the
anticommutator (\ref{car}) will pick up corrective terms, the simplest possible
correction being
\begin{equation} \label{mcar}
\{ \z_{\a}(x^0,\x), \pi^{\b}(x^0,\x') \} = i \eta {\delta_{\a}}^{\b}
\delta(\x-\x') + \ell {{\s^j}_{\a}}^{\b} \p_j \delta(\x-\x').
\end{equation}
with
\begin{equation} \label{ln}
\ell = \ell_P \left( \k + \k' \left( \frac{\ell_P}{{\cal L}}
 \right)^{1+\Upsilon}  \right) \hspace{1cm} \eta = 1 + \tilde \k 
\left( \frac{\ell_P}{{\cal L}} \right)^{1+\Upsilon} + 
\tilde \k ' \left( \frac{\ell_P}{{\cal L}} \right)^{2+2\Upsilon}
\end{equation}
We remark that we may well avoid an effective lagrangian argumentation of
(\ref{mcar}). In the partial expectation value approach of ref \cite{t2},
$H_{eff} = \langle  {\cal L} | H_F |  {\cal L} \rangle$, the possibility of
allowing the deformation (\ref{mcar}) of the canonical structure of the
effective theory is quite natural, since the gravitational sector of the full
Hilbert space affects even the notion of normal order of the matter fields
\cite{t2}. In the alternative approach of refs \cite{amu,alf}, where
$H_{eff} = \langle  {\cal L}, \z | H_F |  {\cal L}, \z \rangle$, the perturbed
dynamics from (\ref{mcar}) would account for the departures of the time
evolution of expectation values of fermion fields from their classical
trajectory, departures that are well known to occur already at order $\hbar$ in
the simplest quantum mechanical systems, even if the initial state is coherent
and properly fine tuned \cite{cs} (This happens because an initial coherent
state evolves into non coherent states. For illustrations in specific 1-D
systems see  e.g. \cite{grandson}.)

Equation (\ref{eh}) together with (\ref{mcar}) and $i \p_t \z_{\g}(t,\x) =
[\z_{\g}(t,\x),H_{eff}]$ give a modified equation for the Weyl spinor. We first
evaluate  the nonzero  commutators:
\begin{eqnarray}
\int d^3x' [\z_{\g}, \pi^{\b}{{\s^k}_{\b}}^{\a} \hat A \p_k \z_{\a}] &=&
\int d^3x' [ i \eta {\delta_{\g}}^{\b} \delta(\x-\x')
+ \ell {{\s^j}_{\g}}^{\b} \p_j \delta(\x-\x')  ] 
{{\s^k}_{\b}}^{\a} \hat A \p'_k \z_{\a} \nonumber \\
&=& i \eta {{\s^k}_{\g}}^{\a} \hat A \p_k \z_{\a} +
\ell  {{\s^j}_{\g}}^{\b} {{\s^k}_{\b}}^{\a} \hat A \p_j \p_k \z_{\a} \nonumber
 \\
&=&  i \eta {{\s^k}_{\g}}^{\a} \hat A \p_k \z_{\a} +
\ell   \hat A \nabla^2 \z_{\g},
\end{eqnarray}
\begin{eqnarray}
\int d^3x' \left[\z_{\g},\frac{i}{2 {\cal L}} \pi^{\a} \hat C' \z_{\a}\right]
 &=&
\frac{i}{2 {\cal L}} \int d^3x' [  i \eta {\delta_{\g}}^{\a} \delta(\x-\x')
+ \ell {{\s^j}_{\g}}^{\a} \p_j \delta(\x-\x') ] \hat C' \z_{\a} \nonumber \\
&=& -\frac{\eta}{2 {\cal L}} \hat C \z_{\g} +\frac{i \ell}{2 {\cal L}} 
 {{\s^j}_{\g}}^{\a} \p_j \hat C \z_{\a}, 
\end{eqnarray}
and 
\begin{multline}
 \int d^3x' [\z_{\g},-\frac{m}{2}\pi^{\b} 
{O'_{\b}}^{\w} \e_{\w \a} \pi^{\a}] = \\
 -\frac{m}{2} \int d^3 x' ( \{ \z_{\g},\pi^{\b}\}  
  {O'_{\b}}^{\w} \e_{\w \a} \pi^{\a} - \pi^{\b}  {{O'_{\b}}^{\w}}
\e_{\w \a} \{\z_{\g},\pi^{\a}\} )   = \\ -i \eta m\a \e_{\g \a} \pi^{\a} -m \ell
\a  {{\s^j}_{\g}}^{\b} \e_{\b \a} \p_j  \pi^{\a} -\eta m \b   {{\s^k}_{\g}}^{\w}
\e_{\w \a} \p_k \pi^{\a} - m \ell \e_{\g \a} \nabla^2 \pi^{\a}
\end{multline}
(here  we have used $ {(\s^{(k} \s^{j)})_{\a}}^{\b}  = \d^{jk} 
{\delta_{\a}}^{\b} $ and  $ (\s^{(k} \e \s^{j)})_{\a \b} = 
-\e_{a \b} \d^{kj}$ and $ {{\s^k}_{\a}}^{\w} \e_{\w \b}={{\s^k}_{\b}}^{\w}
\e_{\w \a}$.)
Collecting terms and suppressing spinor indices we arrive at the following 
equation:
\begin{multline} 
\label{mwe}
\left[i \partial_t -i \eta \hat A \s^j \p_j + \frac{\hat C \eta}{2 {\cal L}}
\right] \z + m \eta \left( \a -i \b \s ^j \p_j \right) i\s^2 \z^* = \\
\ell \left[ \hat A \nabla^2 \z + \frac{i}{2{\cal L}} \hat C  \s ^j \p_j \z 
+  m  (i \a  \s ^j \p_j +\b \nabla^2)  (i \s^2) \z^* \right]
\end{multline}
Setting $\ell=0, \eta=1$ above we recover (\ref{fe}). Note that the operators 
on the right hand side of (\ref{mwe}) are of the same type as those on its left
hand side (with the exception of  higher order corrections like $\nabla \nabla
\z$ in  $\hat A \nabla \z$.) Then there is the possibility that an explicit
evaluation of the $\k$ constants in the effective theory will show that the
Lorentz  symmetry breaking terms on the left and right hand sides of the
equation cancel  each other. In other words, the apparent symmetry breaking
caused by the regularization, normal ordering and by further taking the partial
expectation value of $H_F$ in the gravitational sector, is absorbed by a proper
identification of the conjugate variables in the low energy regime. In fact,
Lorentz symmetry can only be partially  restored, as seen by matching equal
powers of $(\ell_P/{\cal L})$ in  (\ref{mwe}) to show that (\ref{mcar}) and
(\ref{ln}) amount to the following redefinitions of $\hat A, \hat C, \a$ and
$\b$ in  (\ref{fe}):
\begin{eqnarray} \nonumber
\hat A' &=& 1 + \left( \k_1 + \frac{\k \k_4}{2} + \tilde \k
\right) \left( \frac{\ell_P}{{\cal L}} \right)^{1+\Upsilon} 
+ \left(  \k_2 + \frac{\k \k_5}{2} + \frac{\k' \k_4}{2} + \tilde \k \k_1 
+ \tilde \k '\right) \left( \frac{\ell_P}{{\cal L}} \right)^{2+2\Upsilon}  
+ \frac{\k_3}{2} \ell_P ^2 \nabla^2 \\ \nonumber
\hat C' &=& \k_4 \left( \frac{\ell_P}{{\cal L}} \right)^{\Upsilon}  +
(\k_5 +\k_4 \tilde \k) \left( \frac{\ell_P}{{\cal L}} \right)^{1+2\Upsilon}  +
 (\k_6 + \k_5 \tilde \k + \k_4 \tilde \k ')\left( \frac{\ell_P}{{\cal L}}
 \right)^{2+3\Upsilon}   \\
&&+ \left[  \left(\frac{\k_7}{2} - 2(\k \k_1+ \k') \right) 
  \left( \frac{\ell_P}{{\cal L}} \right)^{\Upsilon} 
{\ell_P}^2 - 2 {\cal L} \k \ell_P \right] \nabla^2 \label{mops}\\
\a' &=& 1+ (\k_8+\tilde \k)  \left( \frac{\ell_P}{{\cal L}} \right)^{1+\Upsilon}
\hspace{1cm} \b'/\ell_P = \left( \frac{\k_9}{2} + \k \right) + 
\left(  \frac{\k_{11}}{2} + \k \k_8 + \frac{\tilde \k \k_9}{2} + \k'\right) 
\left( \frac{\ell_P}{{\cal L}} \right)^{1+\Upsilon} \nonumber
\end{eqnarray}
Interestingly enough, this potential restoration of Lorentz symmetry, if only
partial, is enough to solve the contradiction found in \cite{suv}.

\section{Quantum gravity corrections to neutrino propagation}

We analyze the consequences of the modifications introduced by (\ref{mcar}) in
(\ref{mwe}) in the results in \cite{amu}, where the propagation of neutrinos
from  GRB is considered. For these fermions, $p \sim 10^5 GeV$ and  $m \sim
10^{-9} GeV$, thus $m/p \sim p \ell_P \sim 10^{-14} \equiv \e$. Assuming
$\Upsilon \simeq 0$, ${\cal L} \sim 1/p$,  and  keeping  terms up to order
$\e^{2+3 \Upsilon}$, we can see that all terms in  (\ref{mops}) should be kept,
except the $(\frac{\ell_P}{{\cal L}})^{1+\Upsilon}$ piece of  $\b'/\ell_P$.
This is so because $\p_t \sim \nabla \sim \frac{1}{\cal L} \sim p$, whereas $m =
\e p$. Note that the  laplacian term in $\hat C$, which is order $\e^2$, has
picked up a $\k$ correction of order $\e$. This is the most important change
introduced by the correction (\ref{mcar}). It implies helicity dependent effects
of order $p \ell_P$ whenever $\k \neq 0$, as opposed to the predicted 
effects of order $(p \ell_P)^2$ in \cite{amu} (this is seen by noting that the
term $\pm \k_7 (p \ell_p)^ 2/2$ of equation (7) in \cite{amu} gets replaced by $
\mp 2 \k (p \ell_p)$.) Also, the replacements $\k_1 \to \k_1+\k \k_4/2 + \tilde
\k$, etc, may lead to cancellations that switch off some of the predicted QG
effects on neutrinos. None of these predictions can be checked using currently
available experimental data.

\section{Quantum gravity corrections in nuclear physics experiments}

In \cite{suv},  (\ref{dirac}) is applied to non relativistic nucleons. The
weave scale ${\cal L}$ is set equal to $1/m$, $\k_4$ is set equal to zero, and
only leading order corrections are kept. The results are the bounds
(\ref{bounds}). We will instead assume $\Upsilon$ to be a small (less than one)
positive number and use (\ref{mops}) keeping  terms up to  order
$1+ \Upsilon$. Equation  (\ref{mops}) reduces to 
\begin{eqnarray} \nonumber
\hat A' &=& 1 + \left( \k_1 + \tilde \k \right) 
(m \ell_P)^{1+\Upsilon} \hspace{1cm}
\hat C' =  \k_5 (m \ell_P)^{1+2\Upsilon} \\
\a' &=& 1+ ( \k_8 + \tilde \k )  (m\ell_P)^{1+\Upsilon} 
\hspace{1cm} \b'/\ell_P = \left( \frac{\k_9}{2} + \k \right)  \label{mops2}
\end{eqnarray}
and the bounds predicted in \cite{suv} now read
\begin{equation}
\left| \left(\k_1 + \tilde \k \right) (m \ell_P)^{\Upsilon} \right|
< 3 \times 10^{-5} \hspace{1cm} \left| \k_9 + 2\k + \k_5
(m \ell_P)^{2\Upsilon} \right| < 4 \times 10^{-9}
\end{equation}
(the $\k_5$ term should be dropped if $\Upsilon \neq 0$.) These bounds
can safely be satisfied for  $\k$'s of order unity, unless $\Upsilon = \k =
\tilde \k =0$. This shows that  admitting an expansion in non-integer powers of
$(\ell_P/{\cal  L})$, as in \cite{alf}, and the corrections (\ref{car}) to the
anti-commutator  in the semiclassical regime, the contradictions found in
\cite{suv} can be solved without questioning the basic framework of loop quantum
gravity.

\section{Conclusions}

The efforts directed at the construction of a quantum theory of gravity based on
a canonical formalism have led in recent years to a formulation that appears
quite promising, but that is so far incomplete. In particular, a definite
procedure connecting the theory with the low energy regime is still lacking, and
one has to resort to plausibility arguments as to the final form of the theory
in order to make predictions regarding the observable consequences of the theory
in that domain. Nevertheless, using as a guide essentially heuristic notions, a
small number of possible phenomena where the quantum gravity effects might be
sought, together with estimations of their magnitude, have been indicated in the
recent literature. A common feature in all cases, however, is that the
observable data, rather than indicating their presence, seems to impose severe
upper bounds on their possible existence. But, again a common feature of these
effects is they all violate in one form or another (local) Lorentz invariance.
Since it is not at all clear that the complete theory should necessarily lead to
such violation, in this Rapid Communication we have suggested a way in which, in
the fermion sector, a modification of the {\em effective} canonical structure in
the low energy regime might naturally lead to a better agreement with
observational data and compatibility with Lorentz invariance.

\section*{Acknowledgments}

This work was supported in part by grants of the National University of
C\'ordoba, CONICET and Agencia C\'ordoba Ciencia. The authors are researchers of
CONICET (Argentina).

\end{document}